\documentclass[11pt,preprint]{aastex}
\singlespace
\slugcomment{Accepted for the publication in the {\it Astrophysical Journal}: scheduled for the July-20-2004 issue)}
\def\la{\mathrel{\hbox{\rlap{\hbox{\lower4pt\hbox{$\sim$}}}\hbox{$<$}}}}
\def\ga{\mathrel{\hbox{\rlap{\hbox{\lower4pt\hbox{$\sim$}}}\hbox{$>$}}}}

\shortauthors{Park}
\shorttitle{SNR 1987A}

\begin{document}

\title{A {\it Chandra} View of The Morphological And Spectral Evolution 
of Supernova Remnant 1987A}

\author{Sangwook Park}

\affil{Department of Astronomy and Astrophysics,
525 Davey Lab., Pennsylvania State University, University Park, PA. 16802} 
\email{park@astro.psu.edu}

\author{Svetozar A. Zhekov}

\affil{Space Reseach Institute, Moskovska str. 6, Sofia-1000, Bulgaria}

\author{David N. Burrows, Gordon P. Garmire}

\affil{Department of Astronomy and Astrophysics,
525 Davey Lab., Pennsylvania State University, University Park, PA. 16802}

\author{and}

\author{Richard McCray }

\affil{JILA, University of Colorado, Box 440, Boulder, CO. 80309}

\begin{abstract}

We present an update on the results of our monitoring 
observations of the X-ray remnant of supernova (SN) 1987A with the 
{\it Chandra X-Ray Observatory}. As of 2002 December, we have 
performed a total of seven observations of SN 1987A, which allows 
us to monitor the details of the earliest stage of the supernova 
remnant evolution in X-rays. The high angular resolution 
images from the latest data reveal developments of new X-ray bright spots 
in the northwestern and the southwestern portions of the remnant as
well as changes on the eastern side. The observed soft X-ray flux is 
increasing more rapidly than ever, and the latest 0.5$-$2 keV band 
flux ($f_X$ $\sim$ 6 $\times$ 10$^{-13}$ ergs cm$^{-2}$ s$^{-1}$) 
is four times brighter than three years earlier when this monitoring began. 
The overall X-ray emission is primarily from the blast wave shock with 
$kT$ $\sim$ 2.4 keV. As the blast wave approaches the dense circumstellar 
material, the contribution from the decelerated slow shock ($kT$ $\sim$ 
0.22 keV) to the observed X-ray emission is becoming significant. 
The increase of this slow shock contribution over the last two years
is particularly noticeable in the western half of the remnant. 
These results indicate that the shock front is now reaching the main 
body of the inner circumstellar ring and that SN 1987A will be a 
complete ring with dramatic brightening in coming years. Based on 
the best-fit two-shock spectral model, we derive approximate densities 
of the X-ray-emitting regions ($n_e$ $\sim$ 235 cm$^{-3}$ for the fast 
shock and $n_e$ $\sim$ 7500 cm$^{-3}$ for the slow shock). There is no
direct observational evidence to date for a neutron star associated with 
supernova remnant 1987A. We obtain an upper limit on the observed
X-ray luminosity of any embedded point source ($L_X$ $\le$ 1.5 
$\times$ 10$^{34}$ ergs s$^{-1}$) in the 2$-$10 keV band. The X-ray
remnant continues to expand linearly at a rate of 4167 km s$^{-1}$.

\end{abstract}


\section {\label {sec:intro} INTRODUCTION}

Supernova (SN) 1987A, the brightest supernova observed since 1604, 
occurred in the Milky Way's satellite galaxy, the Large Magellanic Cloud 
(LMC). The identification of a blue supergiant (Sanduleak $-$69$^{\circ}$202, 
a B3 I star) as the massive progenitor star \citep{kir87,son87} and the 
detection of the accompanying neutrino burst \citep{koshiba87} indicated 
that SN 1987A was a Type II core-collapse explosion. 
A dramatic brightening (up to three orders of magnitude) over the 
entire electromagnetic spectrum was expected to begin $\sim$15 yr after 
the SN explosion as the shock enters the dense, non-spherically symmetric
circumstellar medium (CSM), the so-called ``inner ring'', which was 
produced by the stellar winds from the red supergiant and the blue 
supergiant phases of the massive progenitor star and then was photoinonized 
by the UV flash from the SN explosion \citep{bur95,lund91,luo91}.  
At an age of $\sim$16 yr (as of 2002 December), optical and X-ray 
observations clearly show that this event has begun \citep{mccray04}, 
signaling the transition of SN 1987A into a supernova remnant (SNR).  
SN/SNR 1987A is an extremely dynamic object, as the fast blast wave shock 
($v$ $\sim$ 4000 km s$^{-1}$) begins to heat the inner ring. 
This remarkable event is the first-ever observation of the ``birth'' 
of a supernova remnant. 

In the early phases 
of SNRs, X-ray observations provide effective shock diagnostics such as 
the electron/ion temperatures, ionization state, electron density, and ISM 
abundances, as well as useful information on the SN nucleosynthesis. With 
its unprecedented high angular resolution of $\sim$0$\farcs$5 in X-rays 
and its moderate spectral resolution (e.g., $E/{\Delta}E$ $\sim$ 8 at $E$ 
= 1 keV for the S3 chip), the Advanced CCD Imaging Spectrometer (ACIS; 
Garmire et al. 2003) on board the {\it Chandra X-Ray Observatory} 
\citep{weiss96} is uniquely suited to monitor the rapid morphological and 
spectral evolution (on the time-scale of months) of the X-ray remnant 
of SN 1987A. 

We have been monitoring SNR 1987A with the {\it Chandra}/ACIS since 
1999 October. As of 2002 December, we have performed a total of seven 
observations (Table~\ref{tbl:tab1}). Results from the first six observations 
have been reported in the literature \citep{bur00,park02,park04,michael02}. 
The sub-arcsecond angular resolution of the {\it Chandra}/ACIS resolved 
the ring-like X-ray morphology of SNR 1987A, which was interpreted as 
X-ray emission from the heated ISM between the forward blast wave and the 
reverse shock \citep{bur00}. The X-ray surface brightness was higher in 
the eastern side, probably due to an asymmetric SN explosion, although 
the origin of such an asymmetric explosion is uncertain. We detected 
X-ray-bright spots in the northeastern and the southeastern portions of 
the X-ray remnant \citep{bur00,park02}. The locations of these X-ray 
spots were roughly coincident with the optically-bright spots. The optical 
bright spots trace where the blast wave shock front cools down and becomes 
radiative as it encounters dense inward protrusions of the inner ring 
\citep{michael00}. In this model, decelerated, yet non-radiative shocks 
near the protrusions may also produce soft X-ray emission. The soft X-ray 
spots positionally coincident with the optical spots were thus consistent 
with the standard model of the blast wave/inner ring interaction. On 
the other hand, the fast shock propagating into the tenuous H{\small II} 
region appeared to produce hard X-ray and radio emission with a smoother 
distribution of surface brightness than those of the soft X-ray and 
optically bright spots \citep{park02}.

The X-ray spectrum was thermal in origin, dominated by line emission 
from the highly ionized elemental species O, Ne, Mg and Si \citep{bur00}. 
The X-ray emitting plasma was described with a plane-parallel shock in 
non-equilibrium ionization (NEI) state with an electron temperature of 
$kT$ $\sim$ 3 keV. A blast wave shock velocity of $v$ $\sim$ 3500 
km s$^{-1}$ was derived from the Doppler broadening of the detected 
X-ray line profiles, which indicated an ion temperature of $kT$ $\sim$ 
17 keV \citep{michael02}. This large difference between measured electron 
and ion temperatures was direct observational evidence for electron-ion 
non-equilibrium behind the shock in this early phase of the SNR. 
The observed X-ray flux was increasing non-linearly as the blast wave 
approached the dense inner ring \citep{park02}. The lightcurve was 
steepening, which indicated the beginning of the predicted dramatic 
brightening of the SNR \citep{park04}. Despite the identification
of a core-collapse SN explosion for SN 1987A, no direct evidence for
the detection of the associated neutron star has been reported. Only 
upper limits on the X-ray luminosity for the embedded pointlike source 
have been estimated \citep{bur00,park02,park04}. We here report on 
results from the latest {\it Chandra} observations of SNR 1987A. 

\section{\label{sec:obs} OBSERVATIONS \& DATA REDUCTION}

The seven {\it Chandra} observations of SNR 1987A are presented in 
Table~\ref{tbl:tab1}. We first screened all data sets with the flight 
timeline filter and turned off the pixel randomization for the highest 
possible angular resolution. We then corrected the spatial and spectral 
degradation of the ACIS data caused by radiation damage, known as 
the charge transfer inefficiency (CTI; Townsley et al. 2000), with the 
methods developed by Townsley et al. (2002a), before further 
standard data screening by status, grade, and energy selections. 
The expected effects of the CTI correction include an increase of 
the number of detected events and improved event energies and energy 
resolution \citep{town00,town02a}. ``Flaring'' pixels were removed and 
{\it ASCA} grades (02346) were selected. Photons between 0.3 keV and 
8.0 keV were extracted for the data analysis. Lightcurves around the 
source region were examined for possible contamination from variable 
background emission and no severe variability was found. The pileup 
fraction was relatively small ($<$10\%) and thus was ignored in the 
analysis. 

We then applied the ``sub-pixel resolution'' method \citep{tsunemi01} 
to improve the angular resolution of the images to better than the CCD 
pixel size. A typical improvement in the angular resolution by 
$\sim$10\% is expected from this method \citep{mori01}. The angular 
size of SNR 1987A is small (the inner ring is only about 1$\farcs$6 
across; e.g., Burrows et al. 1995; Jakobsen et al. 1991), and the ACIS 
detector pixel size (0$\farcs$492) is not adequate to fully resolve 
the remnant. In order to further improve the effective angular resolution 
of the ACIS images, we deconvolved the images using a maximum likelihood 
algorithm \citep{rich72,lucy74} as described in our previous works 
\citep{bur00,park02}. For the image deconvolution, we used 
0$\farcs$125 sky pixels for the data sets with low photon statistics 
($\la$1000 source counts) and 0$\farcs$0625 pixel size for the high 
statistics data sets ($\ga$1000 source counts). 

\section{\label{sec:image} MORPHOLOGICAL EVOLUTION}

\begin{figure}[]
\figurenum{1}
\centerline{\includegraphics[angle=0,width=0.8\textwidth]{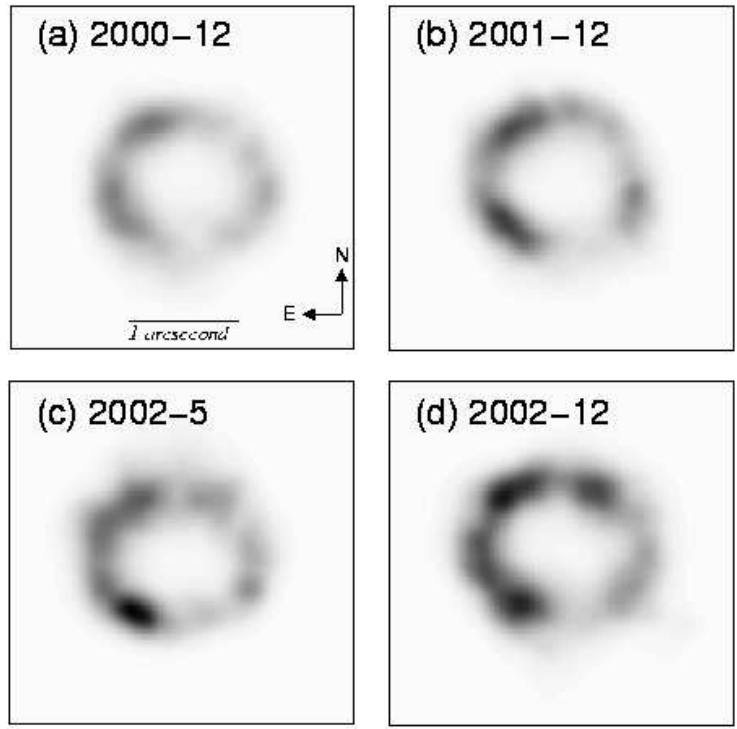}}
\figcaption[]{The gray-scale 0.3$-$8.0 keV broadband images of SNR 1987A.
each image is exposure-corrected and the darker gray-scale is higher 
flux. The image deconvolution has been applied and then the images 
have been smoothed by convolving with a Gaussian ($\sim$0$\farcs$1 FWHM).
\label{fig:fig1}}
\end{figure}

In Figure~\ref{fig:fig1}, we display the ACIS images of SNR 1987A from 
the four observations which have the best photon statistics (ObsID 1967, 
2831, 2832, and 3829; Table~\ref{tbl:tab1}). Images from the other three 
observations with fewer counts have been presented elsewhere 
\citep{bur00,park02}. These broadband X-ray images clearly exhibit the 
dynamical nature of this young SNR. The overall surface brightness has 
doubled over the two-year period ($\sim$0.09 counts s$^{-1}$ in 2000-12 
and $\sim$0.19 counts s$^{-1}$ in 2002-12). While the overall flux 
distribution between the eastern and western sides remains asymmetrical, 
the latest data reveal the development of new X-ray spots in the 
northwest and the southwest. New X-ray-bright spots have also developed 
along the eastern side of the SNR, making the eastern portion of the X-ray 
ring more smooth than before. The overall X-ray morphology of SNR 1987A 
thus appears to become a more complete ring during this two year interval. 
These morphological changes are generally in good agreement with the 
standard model of the blast wave shock front approaching the dense inner 
ring (Michael et al. 2000): i.e., the blast wave reached the northeastern 
side of the inner ring in around 1997 ($\sim$3700 days after the SN 
explosion) when the first optical spot emerged \citep{pun97,garna97}, 
and it is now reaching the western side of the inner ring. As of 2003 
January, the latest optical images taken by the {\it Hubble Space 
Telescope (HST)} revealed that optically bright spots have emerged all 
around the inner ring \citep{mccray04}. This is a good indication of the 
blast wave now encountering the inner ring on the western side, a few 
years after it did on the eastern side. The morphological changes from 
our {\it Chandra} data are in good agreement with those in the optical 
data. We discuss more about this dynamical nature of the blast wave 
in \S~\ref{sec:spectrum}, based on the spectral analysis results.

\begin{figure}[]
\figurenum{2}
\centerline{\includegraphics[angle=0,width=0.8\textwidth]{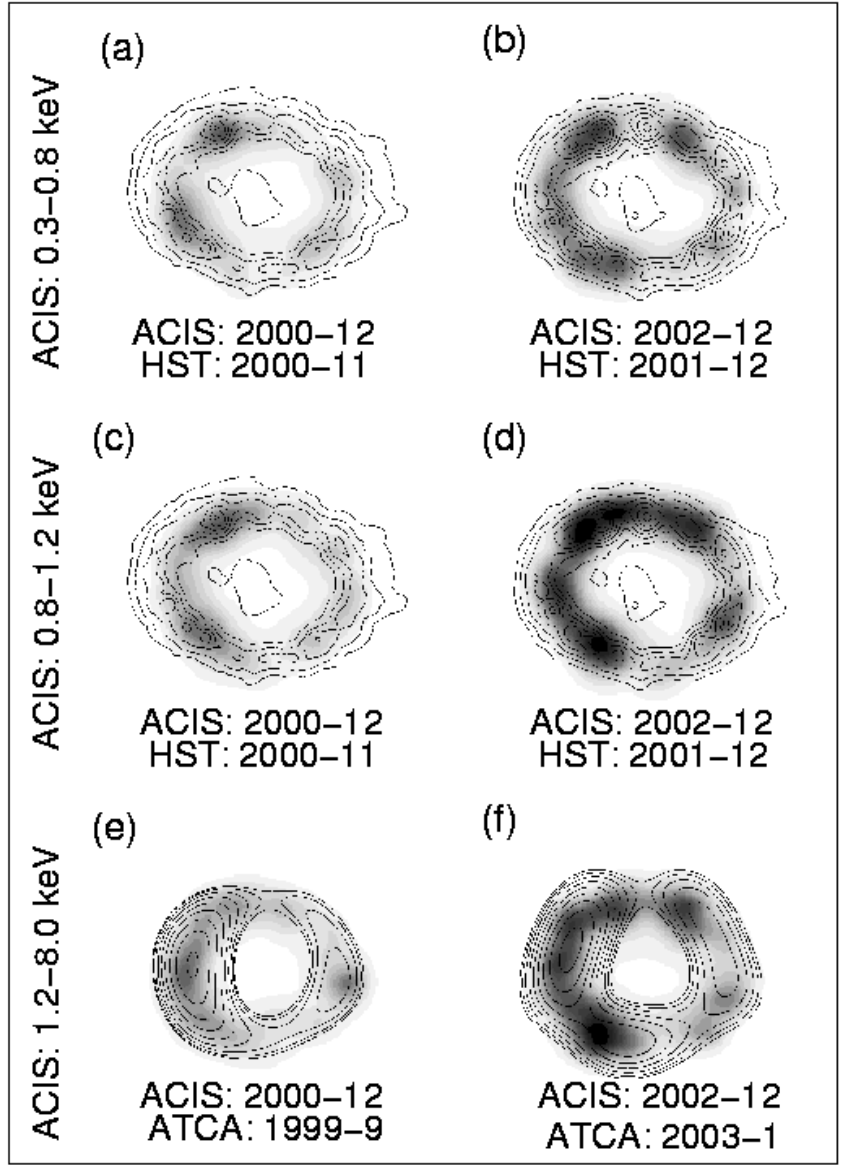}}
\figcaption[]{The broad sub-band ACIS images of SNR 1987A. Each image
has been processed in the same way as those in Figure~\ref{fig:fig1}. 
The {\it HST} contours are overlaid with the O-band (0.3$-$0.8 keV) 
and the Ne-band (0.8-1.2 keV) ACIS images. The 9 GHz radio contours 
taken by Australian Telescope Compact Array ({\it ATCA}) are overlaid
with the H-band (1.2$-$8.0 keV) ACIS images.
\label{fig:fig2}}
\end{figure}

Morphological investigations with broad sub-band images have revealed 
remarkable differences between the soft and the hard band X-ray emission 
features \citep{park02}. The recent observations with significant 
photon statistics allow us to repeat this investigation over a
larger time base. In Figure~\ref{fig:fig2}, we present three 
sub-band images of SNR 1987A for 2000-12 and 2002-12. Both data 
contain comparable numbers of photons ($\ga$9000 counts) so that 
we can make reliable comparisons on the energy-dependent morphological 
changes over this two year interval. The sub-bands were chosen primarily 
to represent the O line features (the 0.3$-$0.8 keV band, ``O-band'' 
hereafter), the Ne line features (the 0.8$-$1.2 keV band, ``Ne-band'' 
hereafer), and the Mg/Si line + any hard-tail emission features (the 
1.2$-$8.0 keV band, ``H-band'' hereafter). The sub-band boundaries
were also chosen to provide comparable photon statistics in each 
sub-band for reliable comparisons among them. In 2000-12, the O- and 
Ne-band X-ray emission was bright only in the northeast and the 
southeast portions of the SNR and these soft X-ray-bright spots were 
positionally coincident with the optically-bright spots 
(Figure~\ref{fig:fig2}a and Figure~\ref{fig:fig2}c). On the other hand, 
the H-band X-ray emission was nearly anti-correlated with the soft X-ray 
features and well correlated with the radio bright features 
(Figure~\ref{fig:fig2}e). This remarkable energy-dependent X-ray 
mophology was interpreted as providing good support for the standard 
model: i.e., the soft X-ray and optical spots are emission from the 
slow shock encountering the dense protrusions of the inner ring and 
the hard X-ray and the radio spots are emission from the fast shock 
propagating into the low density regions in between the dense protrusions 
\citep{park02}. As of 2002-12, the sub-band X-ray images show more 
complex features than in 2000-12 (Figure~\ref{fig:fig2}), and might not 
be adequately described by the simple physical interpretation proposed 
with the 2000-12 data. On the eastern side, the X-ray emission is 
brighter and now appears to be a continuous arc rather than spatially 
separated spots in all three sub-bands (Figure~\ref{fig:fig2}b, 
Figure~\ref{fig:fig2}d and Figure~\ref{fig:fig2}f). On the western side, 
new soft and hard X-ray spots have emerged, particularly in the 
northwestern and the southwestern portions of the inner ring. These 
morphological changes suggest that the blast wave shock front is 
encountering the main part of the inner ring on the eastern side and 
that it has begun reaching the dense protrusions on the western side.

The increase in the surface brightness is more outstanding in the 
Ne- and H-bands compared to the O-band (Figure~\ref{fig:fig2}). The 
count rate ratios between 2002-12 and 2000-12 data, after the quantum 
efficiency (QE) correction (see \S~\ref{sec:spectrum}), are 
1.88$\pm$0.05 in the O-band, 2.08$\pm$0.04 in the Ne-band, and 
2.05$\pm$0.05 in the H-band, while the overall 0.3$-$8.0 keV band count 
rate ratio is 2.06$\pm$0.03. The count rate increase is more significant 
in the Ne-band than in the O-band. These differential photon flux 
increases in the sub-band images suggest spectral changes in the X-ray 
emitting hot plasma over this two year period. 

\begin{figure}[]
\figurenum{3}
\centerline{\includegraphics[angle=0,width=0.8\textwidth]{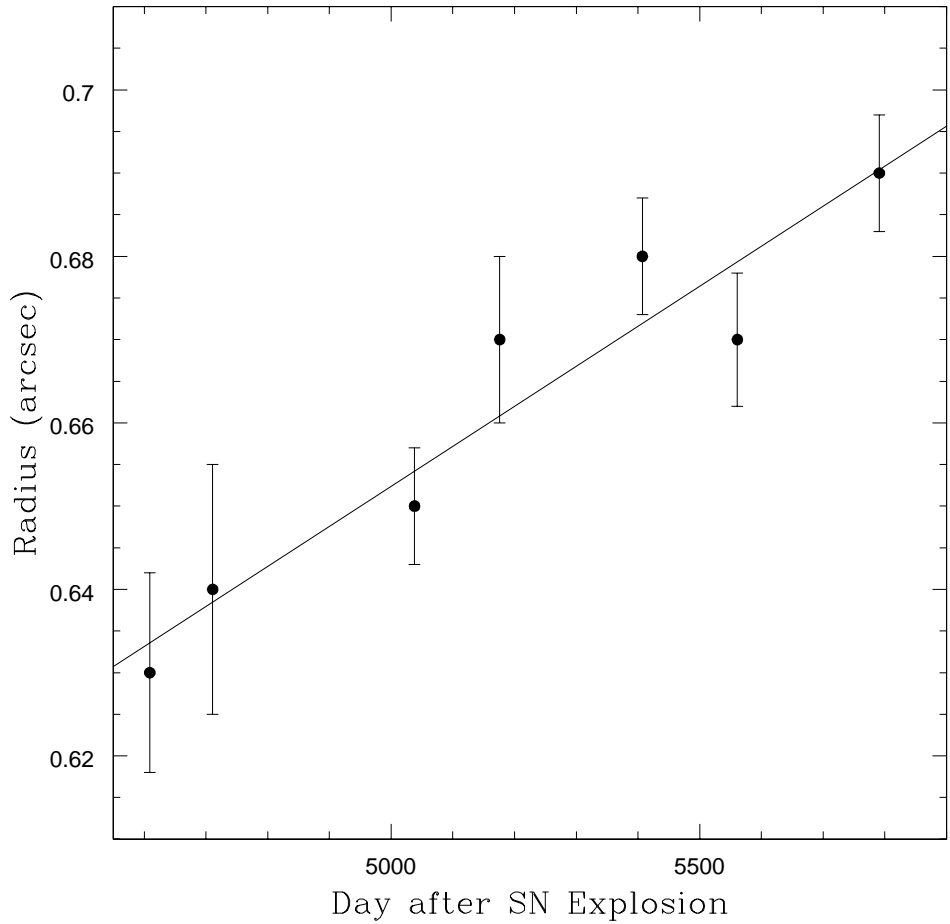}}
\figcaption[]{The long-term variation of the mean radius of the X-ray 
count distribution as obtained with a Gaussian fit. The solid line is 
the best-fit linear increase rate representing an expansion velocity 
of $\sim$4167 km s$^{-1}$.
\label{fig:fig3}}
\end{figure}

Considering the rapid propagation of the supernova blast wave, Park
et al. (2002) investigated the possibility of detecting radial
expansion of the X-ray remnant, and estimated an average angular
expansion of $\sim$0$\farcs$04 between 1999 October and 2001 April.
The measured expansion rate corresponded to a radial velocity of 
5200 $\pm$ 2100 km s$^{-1}$, with a marginal significance 
of $\sim$2.5 $\sigma$. We now have three more data points for a more 
reliable estimate of the radial expansion rate of SNR 1987A. We used 
the same method as described in Park et al. (2002) for this estimate: 
i.e., we constructed radial profiles centered on the mean position of 
the count distribution (or ``center of mass'') with a 0$\farcs$05 
annular bin, and then fitted these radial profiles with Gaussians in 
order to obtain the ``peak'' radius averaged around the X-ray ring. 
The best-fit Gaussian peak radii
versus time are displayed in 
Figure~\ref{fig:fig3}. The peak radius increases by $\sim$0$\farcs$06 
from 1999 October to 2002 December. The best-fit expansion rate of 
SNR 1987A, indicated by the solid line in Figure~\ref{fig:fig3}, is 
4167$\pm$773 km s$^{-1}$, which confirms the previous estimate 
\citep{park02} with higher significance ($\sim$5.4 $\sigma$). 
We note that the X-ray emission is most likely generated between the
forward and the reverse shock, but the bright X-ray features appear 
to be dominated by emission from the forward shock approaching the
dense inner ring. The estimated radii, whose measurements are 
significantly affected by the positions of the bright X-ray-emitting 
features, therefore presumably trace the high emissivity region as 
the blast wave progresses into the dense CSM, rather than measuring 
the actual movement of the X-ray-emitting material. The blast wave shock 
front is likely slowing down as it sweeps through the dense CSM, at least
along the inner ring. Upcoming {\it Chandra} observations will be
monitored for deviation from linear expansion, which would provide
direct evidence for deceleration of the shock front.

\section{\label{sec:spectrum} SPECTRAL EVOLUTION}

The four {\it Chandra}/ACIS observations with good statistics
(ObsID 1967, 2831, 2832, and 3829; Table~\ref{tbl:tab1})  
($\sim$6000$-$9000 counts) allow us to investigate spectral changes 
of the SNR over a period of 24 months.
We extracted the source spectrum of SNR 1987A from 
a circular region with a 2$^{\prime\prime}$$-$3$^{\prime\prime}$ radius 
for each of the four observations. The background was estimated from a 
surrounding annulus with an inner radius of 3$\farcs$5$-$4$^{\prime\prime}$ 
and an outer radius of 7$\farcs$5$-$10$^{\prime\prime}$. Each spectrum has 
been binned to contain a minimum of 50 counts per channel. For the 
spectral analysis of our CTI-corrected data, we have utilized the 
response matrices appropriate for the spectral redistribution of the 
CCD, as generated by Townsley et al. (2002b). The low energy ($E$ $\la$ 
1 keV) QE of the ACIS has degraded because of molecular contamination 
on the optical blocking filter. We corrected this time-dependent QE 
degradation by modifying the ancillary response function for each 
extracted spectrum, utilizing the IDL ACISABS software\footnote{For 
a discussion of this instrumental issue, see 
http://cxc.harvard.edu/cal/Acis/Cal\_prods/qeDeg/index.html.
The software was developed by George Chartas and is available at
http://www.astro.psu.edu/users/chartas/xcontdir/xcont.html.}.
 
\begin{figure}[]
\figurenum{4}
\centerline{\includegraphics[angle=0,width=0.8\textwidth]{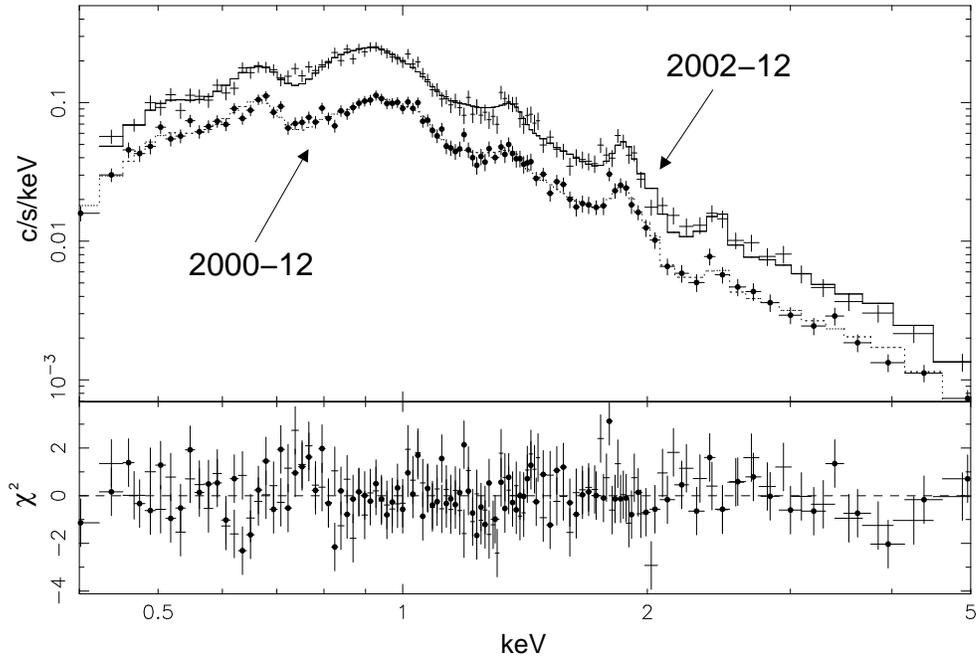}}
\figcaption[]{The X-ray spectrum of SNR 1987A. The best-fit single
NEI model is overlaid for each spectrum taken on 2000-12 and 2002-12.
\label{fig:fig4}}
\end{figure}

\begin{figure}[]
\figurenum{5}
\centerline{\includegraphics[angle=0,width=0.8\textwidth]{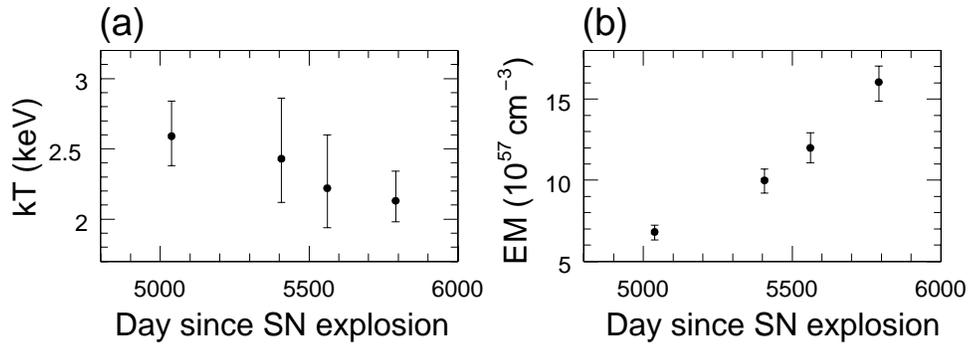}}
\figcaption[]{The electron temperature and the emission measure
variations of SNR 1987A between 2000-12 and 2002-12. 
\label{fig:fig5}}
\end{figure}

We fitted individual X-ray spectra of SNR 1987A with an NEI plane-parallel 
shock model \citep{bor01}, and found that fitted parameters were similar 
among all four spectra. We can therefore obtain the best limits on the 
parameters by performing a combined fit to the four spectra, making the 
assumption that both N$_H$ and the elemental abundances are constant 
during this time interval (the statistics are not good enough to detect 
abundance variations). The spectral contributions from the elements 
He, C, Ca, Ar, and Ni are insignificant in the fitted energy range 
(0.4$-$5.0 keV), so we fixed the abundances of these elements at plausible 
values. He and C were set to the abundances appropriate for the inner 
circumstellar ring \citep{lund96}: He = 2.57 and  C = 0.09, relative to 
solar abundances \citep{anders89}. We fixed the Ca (= 0.34), Ar (= 0.54) 
and Ni (= 0.62) abundances to values appropriate for the LMC ISM 
\citep{russell92} because the ring abundances were unavailable for these 
elemental species in Lundqvist \& Fransson (1996). The elements N, O, Ne, 
Mg, Si, S, and Fe typically have significant effects on the X-ray spectral 
shape in the fitted energy band, and we therefore use the X-ray 
observations to measure these abundances directly in SNRs. In fact, X-ray 
line emission from the most of these species was detected from the 
dispersed spectrum \citep{michael02}, although the poor photon statistics 
did not allow firm measurements of these elemental abundances. Nonetheless, 
earlier ACIS data indicated that fixing these elemental abundances
at the ring or LMC values could not adequately describe the observed X-ray 
spectrum of SNR 1987A \citep{michael02}. We allowed the abundances of these 
elements to vary freely, but constrained them to be the same for all four 
spectra. The temperature and emission measure, on the other hand, were
allowed to vary freely between the four spectra. The best-fit column 
density is $N_H$ = 1.7$^{+0.2}_{-0.3}$ $\times$ 10$^{21}$ cm$^{-2}$ 
(2$\sigma$ uncertainties are quoted throughout this paper). The results 
of this spectral fitting are summarized in Table~\ref{tbl:tab2} and 
Table~\ref{tbl:tab3}. In Figure~\ref{fig:fig4}, we present the X-ray 
spectra of SNR 1987A from data taken on 2000-12 and 2002-12. The best-fit 
NEI shock model is overlaid onto each spectrum. Figure~\ref{fig:fig4} 
shows that the overall spectral shape has not changed significantly
as the X-ray flux increased over the past two years. While the temporal 
variation of the ionization timescale is uncertain with the current data, 
the increase in the emission measure is substantial (Table~\ref{tbl:tab3}). 
There also appears to be a monotonic decrease in the electron temperature 
for the last two years (Table~\ref{tbl:tab3}). Figure~\ref{fig:fig5} 
displays these temporal variations of the electron temperature and the 
volume emission measure. These overall variations of the electron 
temperature and the emission measure are typical signatures of a shock 
slowing down by encountering dense material, which is in good agreement 
with the standard model of the blast wave-inner ring interaction for 
SNR 1987A.

\begin{figure}[]
\figurenum{6}
\centerline{\includegraphics[angle=0,width=0.8\textwidth]{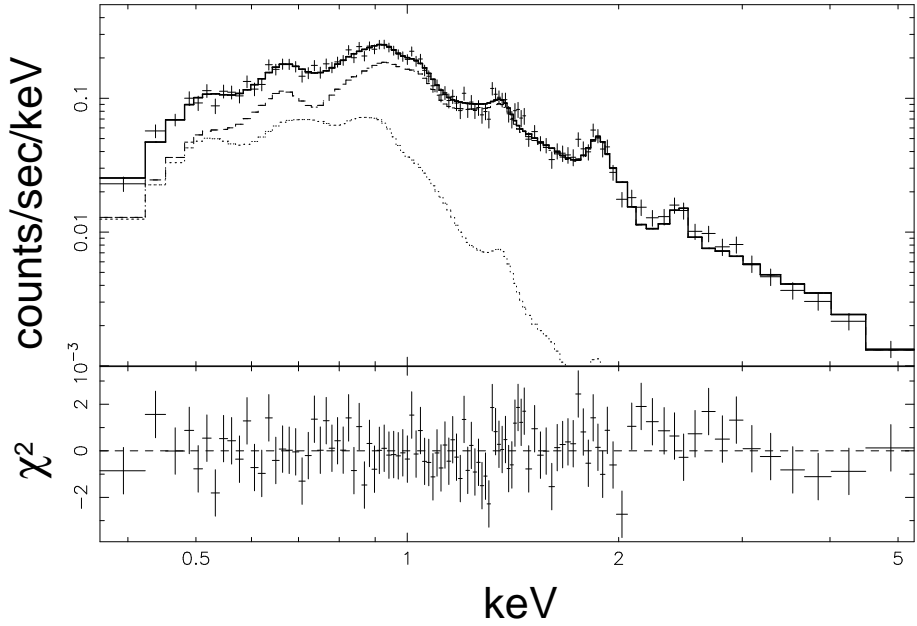}}
\figcaption[]{The X-ray spectrum of SNR 1987A as of 2002-12. The best-fit
two-temperature model is overlaid.
\label{fig:fig6}}
\end{figure}

As the blast wave begins to interact with the dense inner ring, the
X-ray spectrum is expected to show evidence for increasing contributions
from slow shocks ($v$ $\sim$ 500 km s$^{-1}$) \citep{michael02}. 
In fact, our spectral analysis with a single temperature model suggested 
a decrease of the best-fit electron temperature over the past two years. 
We thus examined the slow shock contribution by fitting the latest 
spectrum (2002-12) with a two-temperature model. We forced elemental 
abundances and the foreground column to be the same for both the soft 
and hard components, while allowing the electron temperature, ionization 
timescale, and emission measure to vary separately for each component. 
The fitted abundances are generally consistent with those obtained above. 
We found that the soft component indicated a highly 
advanced ionization state ($n_et$ $\sim$ 10$^{13}$ cm$^{-3}$ s), and 
can be described by a thermal plasma in collisional ionization equilibrium 
(CIE). We thus repeated the spectral fitting with a CIE soft component + 
an NEI hard component by fixing elemental abundances at the best-fit 
values as presented in Table~\ref{tbl:tab2} (Figure~\ref{fig:fig6}). 
The best-fit parameters from this two-component model fit for the 
2002-12 data are presented in Table~\ref{tbl:tab4}. 
While the implemention of this two-temperature model was physically 
motivated by considering the shock-CSM interaction, the presence of
the additional soft component appears to be supported by the statistics
as well: i.e., we separately fitted the 2002-12 data with a single 
temperature NEI model ($\chi^2$/$\nu$ = 115.84/101), and an F-test 
suggested that the statistical improvement in the fit with the 
two-temperature model is significant (F-probability is $\sim$4 $\times$ 
10$^{-5}$). The best-fit foreground absorption ($N_H$ $\sim$ 2.3 
$\times$ 10$^{21}$ cm$^{-2}$) is some $\sim$30\% higher than that from 
the single-temperature model fit ($N_H$ $\sim$ 1.7 $\times$ 10$^{21}$ 
cm$^{-2}$), which is perhaps expected with the addition of the soft 
component. The higher absorption is in better agreement with 
that measured by UV spectroscopy ($N_H$ $\sim$ 3 $\times$ 10$^{21}$ 
cm$^{-2}$; e.g., Fitzpatrick \& Walborn 1990), and is presumably more 
realistic. The low electron temperature ($kT$ = 0.22 keV) for the soft 
component implies a significantly decelerated shock velocity of $v$ 
$\sim$ 400 km s$^{-1}$ (assuming an electron-ion equilibrium for this 
component, as supported by its CIE state). The X-ray flux from 
the slow shock provides $\sim$18\% of the total flux in the 0.5$-$10 
keV band.

Based on the measured $EM$s (Table~\ref{tbl:tab4}), it is evident 
that the electron density for the soft component is higher than
that for the hard component, even assuming the ``same'' X-ray-emitting 
volume for each component. We can estimate the densities for the 
X-ray-emitting regions by considering reasonable geometries for the 
emission volumes. For the fast shock component ($kT$ = 2.44 keV), we 
assume a simple geometry of a spherical shell with an inner radius of 
0$\farcs$6 (i.e., $\sim$4.5 $\times$ 10$^{17}$ cm at the LMC distance of 
50 kpc) where the blast wave began to enter the H{\small II} region at 
an age of $\sim$1200 days \citep{man02}. Assuming the blast wave has 
been propagating at a constant velocity of $\sim$3000 km s$^{-1}$\cite{man02} 
since then, the outer radius is 0$\farcs$76 or $\sim$5.7 $\times$ 10$^{17}$ 
cm as of 2002-12 (5791 days). We also assume $n_e$ $\approx$ 1.5 $n_H$ 
for the ring abundances (e.g., Masai \& Nomoto 1994). An electron density 
of $n_e$ $\sim$ 235 cm$^{-3}$ is then derived from the best-fit emission 
measure of the hard component (Table~\ref{tbl:tab4}). The density of the 
dense CSM that slows down the blast wave to produce the bulk of the soft 
X-ray emission is more difficult to estimate because this dense region, 
containing the inward protrusions, is not resolved and thus the geometry of 
the X-ray-emitting volume is uncertain. Since the optical spots are now 
present all around the inner ring and the soft X-ray emission appears to 
be more continuous along the X-ray ring than ever, we make a simple 
assumption of a shell-like soft X-ray-emitting region. This soft X-ray 
emission is most likely from along the inner ring, whose thickness is 
$\sim$0$\farcs$088 ($\sim$0.02 pc at the distance of 50 kpc; Jakobsen 
et al. 1991). Considering this ring-like geometry, we may thus use a 
cylindrical shell with the ring thickness for the soft component instead 
of a spherical shell. We assumed a fast shock velocity of 3000~km s$^{-1}$ 
between days 1200 and 3700, when the shock began interacting with the CSM 
and slowed down to 400~kms$^{-1}$. With these assumptions, the inner 
radius of the cylindrical shell is $\sim$5.0 $\times$ 10$^{17}$ cm and 
its outer radius is $\sim$5.1 $\times$ 10$^{17}$ cm. The estimated average 
thickness of the dense CSM ($\sim$10$^{16}$ cm) is consistent with the 
``characteristic dimension'' of optical spot 1 \citep{michael00}. With 
this geometry for the X-ray-emitting region, the best-fit volume 
emission measure for the soft component (Table~\ref{tbl:tab4}) implies 
an electron density of $n_e$ $\sim$ 7500 cm$^{-3}$. Previous estimates 
of the densities for the H{\small II} region and the inner ring are on 
the order of $\sim$10$^2$ cm$^{-3}$ and $\sim$10$^4$ cm$^{-3}$, 
respectively \citep{chev95,bor97,lund96}. Considering the systematic 
uncertainties in the adopted geometries and the fact that the blast wave 
might not have encountered the densest portion of the inner ring yet, 
the derived densities are in good agreement with the previous results.

The latest X-ray images reveal the development of new X-ray spots
on the western side of the SNR, which suggests that the blast wave is
approaching the inner ring in the west as well as in the east
(\S~\ref{sec:image}). In order to quantitatively investigate this 
hypothesis, we performed two-component model fitting for the 
eastern and western halves of the SNR, and compared the results 
between 2000-12 and 2002-12. Because of the small angular size and 
the apparent asymmetric flux distribution of the SNR, defining
the ``eastern'' and the ``western'' halves is dependent on how we  
determine the center of the SNR (e.g., the ``center of mass'' of the 
intensity distribution vs. the geometrical center; Michael et al. 2002). 
Our purpose in performing this spectral analysis by halves is to compare 
the flux variations between 2000-12 and 2002-12, presumably caused by 
the emergence of the new X-ray spots in the western side. For this
purpose, we chose centers close to the positions of the faintest 
X-ray flux near the geometrical center of the remnant based on the 
broadband images. The eastern and western halves of the SNR, determined 
by these ``centers'', are intended to effectively separate the bright 
X-ray spots in the northeastern and northwestern sides of the X-ray ring. 
The best-fit spectral parameters for the two halves are generally 
consistent with those from the entire remnant, although less constrained 
due to the reduced photon statistics. The X-ray flux variations between 
the eastern and western sides of SNR 1987A for the last two years are 
presented in Table~\ref{tbl:tab5}. We found that the 0.5$-$10 keV band 
X-ray flux from the slow shock (the soft spectral component) has 
marginally increased by $\sim$35\% in the eastern half between 2000 and 
2002. By contrast, the X-ray flux from the slow shock in the western 
half has significantly increased by a factor of more than 5 for the same 
time period. For comparison, the observed X-ray flux originating from 
the fast shock (the hard spectral component) has only increased by a 
factor of $\sim$2 in both the eastern and the western halves. These 
differential flux changes between the eastern and the western sides of 
the SNR indicate that the blast wave has reached the dense inner ring 
and is slowing down in the western side of the SNR some $\sim$5 years 
after it did in the eastern side when the first optical spot emerged. 
 
\section{\label{sec:lc} X-ray Lightcurve}

\begin{figure}[]
\figurenum{7}
\centerline{\includegraphics[angle=0,width=0.8\textwidth]{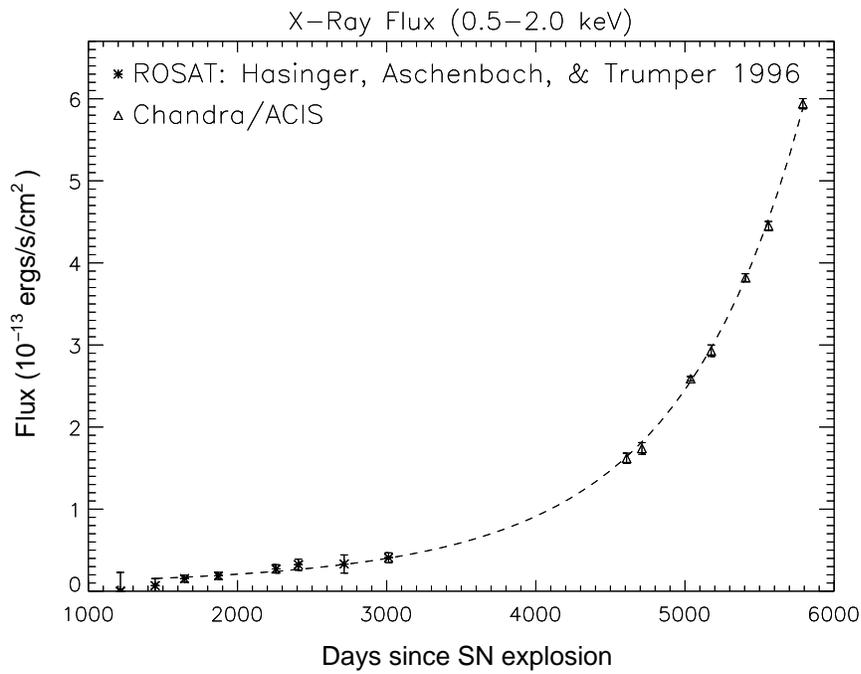}}
\figcaption[]{The X-ray light curve of SNR 1987A. The best-fit
model with an exponential density distribution is overlaid
with a dashed curve.
\label{fig:fig7}}
\end{figure}

A non-linear increase of the X-ray flux from SNR 1987A over the
last decade has been reported \citep{park02,park04}. We present the 
latest X-ray lightcurve of SNR 1987A by combining the {\it ROSAT} and 
{\it Chandra} data (Table~\ref{tbl:tab6}; Figure~\ref{fig:fig7}). 
The {\it ROSAT} data were obtained from 
Hasinger, Aschenbach, \& Tr\"umper (1996). The first four {\it Chandra} 
fluxes have been updated by correcting the QE for contamination, 
which had not been characterized at the time of Park et al. (2002).
The QE correction resulted in some $\sim$6\%$-$20\% increase 
in the estimated X-ray fluxes for those four {\it Chandra} observations 
\citep{park04}. With the earlier observations, only single-temperature 
models were feasible because of the poor photon statistics.
For consistency, we present the soft X-ray fluxes estimated from the 
single-temperature NEI model for all seven {\it Chandra} observations 
in Table~\ref{tbl:tab6} and Figure~\ref{fig:fig7}. As of 2002-12, the 
0.5$-$2 keV band X-ray flux from SNR 1987A is $f_X$ $\sim$ 5.95 $\times$ 
10$^{-13}$ ergs cm$^{-2}$ s$^{-1}$, which is nearly four times brighter 
than 1999-10 when we began this monitoring observations with {\it Chandra}. 

The X-ray flux increase rate has apparently been steepening, and 
cannot be fitted with simple linear or quadratic functions. We thus 
attempted more complex non-linear models to fit the lightcurve. 
Here we present a simple model of a blast wave propagating into an
exponentially increasing density distribution. The X-ray flux 
can be expressed as $f_X$ $\propto$ $n^2_eVT^{-0.6}$ for 0.01 keV 
$<~kT~<$ 3.4 keV, where $V$ is the X-ray emission volume (e.g., 
McKee \& Cowie 1977). We assumed $V$ $\propto$ $R^3$ 
(where $R$ is the radial distance from the center of the SNR) and 
a constant $T$. We note that the electron temperature appears to be 
decreasing during this period (\S~\ref{sec:spectrum}),
but this decrease is small and we assume a constant temperature  
for the purpose of this simple fit. 
(The measured temperature change between 2000-12 and 
2002-12 would imply only $\sim$10\% change in the X-ray flux estimate.)
The blast wave propagates through a low-denisty H{\small II} region 
before encountering the high-density inner ring; thus we expect an 
intermediate density region at the transition between the H{\small II} 
region and the inner ring. We assumed an exponential density 
distribution along the radius of the SNR in this transition region. 
The X-ray flux is then
\begin{equation}
f_X \propto R^3(n_0 + n_re^{-\frac{R_r-R}{D}})^2 ,
\label{eq:eq1}
\end{equation} 
where $n_0$ is the electron density in the H{\small II} region,
$R_r$ is the radius of the inner ring, $n_r$ is the electron density 
of the inner ring at $R = R_r$, and $D$ is a characteristic {\it 
scale height} of the exponential density distribution. Now, we define 
$R_0$ as the radius at which the blast wave started to interact 
with the H{\small II} region, $t$ as the time since the SN explosion, 
$v$ as the shock velocity, and $t_0$ as $t$ at $R$ = $R_0$. 
We can then rewrite equation (1) as
\begin{equation}
f_X = f_0[1 + \beta(\tau-1)]^3[1 + 
qe^{-\frac{\alpha-1-\beta(\tau-1)}{s}}]^2 ,
\label{eq:eq2}
\end{equation}
where $f_0$ is the X-ray flux at $t$ = $t_0$, $\tau$ $\equiv$ 
$t\over{t_0}$, $q$ is the density ratio between the inner ring and the 
H{\small II} region $n_r\over{n_0}$, $s$ $\equiv$ $D\over{R_0}$ is a 
modified characteristic scale, $\alpha$ is the radius ratio of 
$R_r\over{R_0}$, and $\beta$ $\equiv$ $\frac{v~t_0}{R_0}$. Based on 
the radio data \citep{man02}, we chose $R_0$ = 0$\farcs$6, $v$ = 3000 
km s$^{-1}$, and $t_0$ = 1200 days since the SN explosion. We adopted 
$R_r$ = 0$\farcs$83 from the optical data (e.g., Jakobsen et al. 1991). 
We can therefore fit the X-ray lightcurve in terms of $f_0$, $q$, and $s$. 
The best-fit parameters are $f_0$ = (8.0$\pm$2.0) $\times$ 10$^{-15}$ 
ergs cm$^{-2}$ s$^{-1}$, $n_r\over{n_0}$ = 17.6$\pm$1.9, and $D$ = 
0$\farcs$057$\pm$0$\farcs$003 with $\chi^2/\nu$ = 7.4/11. This best-fit 
model is displayed as a dashed-curve in Figure~\ref{fig:fig7}. 
The fitted $f_0$ is in good agreement with the {\it ROSAT} data. The 
best-fit density ratio is considerably lower than the estimates from the 
modeling of previous observations in other wavelengths (typically
$n_r\over{n_0}$ $\sim$ 100; e.g., Chevalier \& Dwarkadas 1995;
Borkowski, Blondin, \& McCray 1997; Lundqvist \& Franssen 1996).
While it is possible that this is due to the simplicity of our model,
we note that this density ratio is also consistent with that derived from 
our spectral analysis (\S~\ref{sec:spectrum}) within a factor of $\sim$2.

The data points used in our fit are {\it incomplete}: 
i.e., the blast wave shock front has not yet swept through the 
densest portions of the ring, hence our X-ray lightcurve has not 
yet peaked.  We thus expect even steeper flux increases with upcoming 
data. 

\section{\label{sec:ns} Neutron Star}

The identification of SN 1987A as a core-collapse SN explosion,
as evidenced by the massive, blue supergiant progenitor star and by 
the detection of the neutrino burst coincident with the SN explosion, 
predicts the existence of a compact remnant, most likely a neutron star, 
at the center of the SNR. However, we have not detected X-ray emission 
from the embedded central point source. This non-detection of the 
central point source is not surprising because the stellar debris from 
the SN explosion is expected to remain opaque in the radio through X-ray 
wavelengths for decades after the SN event \citep{fran87}. We have  
estimated upper limits for the X-ray emission from the embedded point 
source by simulating a point source at the center of the observed X-ray 
images of SNR 1987A using the Monte Carlo technique described previously
\citep{bur00,park02}. We obtain a 90\% confidence upper limit for the 
unabsorbed X-ray luminosity, $L_X$ = 1.5 $\times$ 10$^{34}$ ergs s$^{-1}$,
in the 2$-$10 keV band for any embedded central point source, somewhat 
lower than previous limits.

\section{\label{sec:sum} Summary \& Future}

As of 2002 December, we have performed a total of seven observations
of SNR 1987A with the {\it Chandra}/ACIS. With superb angular 
resolution and moderate spectral resolution, the {\it Chandra}/ACIS 
data have provided a unique opportunity to monitor the morphological 
and spectral evolution of the X-ray remnant of SN 1987A. The X-ray 
remnant is ring-like with an asymmetric flux distribution between 
the eastern and western sides. The overall broadband surface brightness 
has doubled between 2000 December and 2002 December. We detected the 
development of new X-ray-bright spots in the western side of the 
remnant in addition to the previously-known features in the eastern side. 
The X-ray emission features in the eastern side are now more continuous 
than before, while new X-ray spots are emerging in the western side. 
These morphological changes appear to indicate that the blast wave is 
approaching closer to the main portion of the dense inner ring. 
We detect an average angular expansion of $\sim$0$\farcs$06 between 
1999 October and 2002 December, which corresponds to an average radial 
velocity of $v$ $\sim$ 4167 km s$^{-1}$ for the last four years.

Our spectral analysis suggests that the average electron temperature 
of the shock has been decreasing, whereas the observed flux significantly 
increased. The best-fit elemental abundances are generally consistent 
with those of the LMC ISM, and thus the X-ray emission in SNR 1987A 
appears to be dominated by that from the shocked ISM rather than shocked 
metal-rich ejecta. A two-temperature shock model indicates that the 
contribution from the decelerated (by the dense CSM) slow shock 
($v$ $\sim$ 400 km s$^{-1}$) to the observed X-ray flux is $\sim$18\% 
of the total flux as of 2002 December. It is notable that the slow shock 
contribution has increased more significantly in the western side than 
in the eastern side since 2000 December. This spectral evolution is 
consistent with the morphological evolution. Both imply that the blast 
wave is now encountering the dense circumstellar material in the western 
side in addition to the eastern side. Based on the two-temperature model, 
we derive electron densities $n_e$ $\sim$ 235 cm$^{-3}$ for the
hard X-ray emission ($kT$ = 2.44 keV) from the fast shock and 
$n_e$ $\sim$ 7500 cm$^{-3}$ for the soft component ($kT$ = 0.22 keV) 
from the slow shock decelerated by the dense inner ring.

The X-ray lightcurve has been steepening rapidly since 2000.
As of 2002 December, the 0.5$-$2 keV band X-ray flux ($f_X$ $\sim$
6 $\times$ 10$^{-13}$ ergs cm$^{-2}$ s$^{-1}$) is nearly four times
brighter than in 1999 October when this monitoring program began.
We fit the X-ray lightcurve with a non-linear model by assuming an 
exponential density distribution in the transition regions between 
the H{\small II} region and the inner ring. The best-fit model implies
a density contrast of $\sim$17 between these two regions, which is consistent
with the density estimates from the spectral fitting. This result
is also in plausible agreement with the previous density estimates.

Our results clearly demonstrate the extremely dynamic nature of 
SNR 1987A. The morphological and spectral characteristics are rapidly 
changing on a time scale of months. Moreover, the latest
information obtained by our {\it Chandra} data strongly indicates
that the predicted dramatic brightening (up to nearly three orders
of magnitude), as the blast wave sweeps through the dense CSM,
might be underway. This remarkable event represents the first-ever 
observation of the {\it birth of a supernova remnant}. This monitoring 
program of SNR 1987A should thus continue in coming years. In addition
to the imaging observations, high spectral resolution grating 
observations with {\it Chandra} and {\it XMM-Newton} should be
performed in order to determine the details of the shock parameters
such as the electron/ion temperatures, ionization states, the elemental
abundances, and the 3-dimensional structure of the blast wave shock
and the ejecta material.

Although the creation of a compact remnant, probably a neutron star,
is expected from the identified core-collapse explosion for SN 1987A, 
we have yet to detect direct observational evidence for a central 
pointlike source. This non-detection of a point source appears to be 
caused by the cold, dense stellar debris around the center of the SNR, 
which will most likely remain optically thick in the X-ray band for 
more than a decade. We obtain an upper limit on the 
observed 2$-$10 keV X-ray luminosity from any embedded central point 
source of $L_X$ $\sim$ 1.5 $\times$ 10$^{34}$ ergs s$^{-1}$.

\acknowledgments

The authors thank P. Challis and the Supernova INtensive Study (SINS)
collaboration for providing their {\it HST} images. We also thank
D. Manchester and B. Gaensler for providing the radio images taken 
with {\it ATCA}. This work was supported in part by NASA under
contract NAG8-01128 and by SAO under grant GO1-2064b and GO2-3098a.

\clearpage

\begin{deluxetable}{ccccc}
\footnotesize
\tablecaption{{\it Chandra}/ACIS Observations of SNR 1987A
\label{tbl:tab1}}
\tablewidth{0pt}
\tablehead{\colhead{Observation ID} & \colhead{Date (Age\tablenotemark{a}~)} & 
\colhead{Instrument} & \colhead{Exposure (ks)} & \colhead{Source Counts}}

\startdata

124+1387\tablenotemark{b} & 1999-10-06 (4609) & ACIS-S + HETG & 116 & 690 \\
122 & 2000-01-17 (4711) & ACIS-S3 & 9 & 607 \\
1967 & 2000-12-07 (5038) & ACIS-S3 & 99 & 9031 \\
1044 & 2001-04-25 (5176) & ACIS-S3 & 18 & 1800 \\
2831 & 2001-12-12 (5407) & ACIS-S3 & 49 & 6226 \\
2832 & 2002-05-15 (5561) & ACIS-S3 & 44 & 6429 \\
3829 & 2002-12-31 (5791) & ACIS-S3 & 49 & 9274  \\
\enddata

\tablenotetext{a}{Day since the SN explosion.}
\tablenotetext{b}{The first observation was split into two sequences, which
were combined in the analysis.} 
\end{deluxetable}

\begin{deluxetable}{cccc}
\footnotesize
\tablecaption{Best-fit Elemental Abundances.
\label{tbl:tab2}}
\tablewidth{0pt}
\tablehead{\colhead{Element} & \colhead{Abundance\tablenotemark{a}} & 
\colhead{Element} & \colhead{Abundance\tablenotemark{a}}}

\startdata

\vspace{1.0mm}
 He & 2.57 (fixed) & Si & 0.32$^{+0.07}_{-0.06}$ \\
\vspace{1.0mm}
 C  & 0.09 (fixed) & S  & 0.76$^{+0.24}_{-0.23}$ \\
\vspace{1.0mm}
 N  & 0.45$^{+0.10}_{-0.09}$ & Ca & 0.34 (fixed) \\
\vspace{1.0mm}
 O  & 0.10$^{+0.02}_{-0.01}$ & Ar & 0.54 (fixed) \\
\vspace{1.0mm}
 Ne & 0.22$^{+0.03}_{-0.03}$ & Fe & 0.13$^{+0.02}_{-0.02}$ \\ 
\vspace{1.0mm}
 Mg & 0.16$^{+0.04}_{-0.04}$ & Ni & 0.62 (fixed) \\
\enddata

\tablenotetext{a}{Abundances with respect to Solar.}
\end{deluxetable}

\begin{deluxetable}{lccc}
\footnotesize
\tablecaption{Best-fit Shock Parameters\tablenotemark{a}
\label{tbl:tab3}}
\tablewidth{0pt}
\tablehead{\colhead{Date} & \colhead{Electron Temperature} & \colhead{Ionization 
Timescale} & \colhead{Emission Measure} \\
     & \colhead{(keV)} & \colhead{(10$^{10}$ cm$^{-3}$ s)} & 
\colhead{(10$^{57}$ cm$^{-3}$)}}

\startdata

\vspace{1.0mm}
2000-12-7 & 2.59$^{+0.25}_{-0.21}$ & 3.27$^{+0.66}_{-0.50}$ & 
6.81$^{+0.42}_{-0.48}$ \\
\vspace{1.0mm}
2001-12-12 & 2.43$^{+0.43}_{-0.31}$ & 3.04$^{+0.70}_{-0.52}$ & 
9.99$^{+0.69}_{-0.78}$ \\
\vspace{1.0mm}
2002-5-15  & 2.22$^{+0.38}_{-0.28}$ & 3.46$^{+0.82}_{-0.60}$ & 
12.00$^{+0.93}_{-0.93}$ \\
\vspace{1.0mm}
2002-12-31 & 2.13$^{+0.21}_{-0.15}$ & 3.38$^{+0.63}_{-0.50}$ & 
16.05$^{+0.99}_{-1.17}$ \\
\enddata

\tablenotetext{a}{$\chi^2$/$\nu$ = 359.0/352}
\end{deluxetable}

\begin{deluxetable}{lcc}
\footnotesize
\tablecaption{Results from the Two-Temperature Model Fit as of 
2002-12\tablenotemark{a}
\label{tbl:tab4}}
\tablewidth{0pt}
\tablehead{\colhead{Parameters} & \colhead{Soft Component (CIE)} & 
\colhead{Hard Component (NEI)}}

\startdata 

\vspace{1.0mm}
 $N_H$ (10$^{21}$ cm$^{-2}$) & 2.3$^{+0.6}_{-0.5}$ & 2.3$^{+0.6}_{-0.5}$ \\
\vspace{1.0mm}
 $kT_e$ (keV) & 0.22$^{+0.05}_{-0.02}$ & 2.44$^{+0.27}_{-0.23}$ \\
\vspace{1.0mm}
 $n_et$ (10$^{10}$ cm$^{-3}$ s) & - & 4.38$^{+1.42}_{-0.88}$ \\
\vspace{1.0mm}
 $EM$\tablenotemark{b}~(10$^{57}$ cm$^{-3}$) & 59.78$^{+10.71}_{-10.23}$ & 
14.25$^{+0.60}_{-0.57}$ \\
\vspace{1.0mm}
 $f_X$\tablenotemark{c} (10$^{-13}$ ergs cm$^{-2}$ s$^{-1}$) & 
1.57$^{+0.28}_{-0.27}$ & 7.22$^{+0.30}_{-0.29}$ \\
\vspace{1.0mm}
 $L_X$\tablenotemark{c} (10$^{35}$ ergs s$^{-1}$) & 1.61 & 3.72 \\
\enddata

\tablenotetext{a}{$\chi^2$/$\nu$ = 94.29/99.}
\tablenotetext{b}{The uncertainties were estimated after fixing $N_H$ and 
$kT$ at the best-fit values.}
\tablenotetext{c}{The X-ray fluxes and luminosities were estimated in the 
0.5$-$10.0 keV band.}

\end{deluxetable}

\begin{deluxetable}{ccccc}
\footnotesize
\tablecaption{X-ray Fluxes for the Eastern and Western sides of 
SNR 1987A.
\label{tbl:tab5}}
\tablewidth{0pt}
\tablehead{ & \multicolumn{2}{c}{2000-12} & \multicolumn{2}{c}{2002-12} \\
& East & West & East & West }

\startdata 

\vspace{1.0mm}
 Soft Component\tablenotemark{a} & 0.43$^{+0.07}_{-0.08}$ & 
0.17$^{+0.04}_{-0.04}$ & 0.58$^{+0.25}_{-0.25}$ & 0.92$^{+0.16}_{-0.19}$\\
\vspace{1.0mm}
 Hard Component\tablenotemark{a}  & 1.92$^{+0.16}_{-0.13}$ & 
1.46$^{+0.09}_{-0.11}$ & 4.07$^{+0.43}_{-0.43}$ & 3.10$^{+0.33}_{-0.26}$ \\
%$\chi^2$/$\nu$ & 106.9/106 & & 90.1/89 & & 91.3/99 & & 92.5/94 \\
\enddata

\tablenotetext{a}{The 0.5$-$10 keV band fluxes in units of 
10$^{-13}$ ergs cm$^{-2}$ s$^{-1}$. The uncertainties were estimated 
after fixing $N_H$ and abundances at the best-fit values presented
in Table~\ref{tbl:tab2} and Table~\ref{tbl:tab4}.}

\end{deluxetable}

\begin{deluxetable}{cccc}
\footnotesize
\tablecaption{The 0.5$-$2.0 keV band flux of SNR 1987A.
\label{tbl:tab6}}
\tablewidth{0pt}
\tablehead{\colhead{Age\tablenotemark{a}} & \colhead{Observed Flux ({\it ROSAT})}
& \colhead{Age\tablenotemark{a}} & \colhead{Observed Flux ({\it Chandra})}\\ 
\colhead{(days)} & \colhead{(10$^{-13}$ ergs cm$^{-2}$ s$^{-1}$)} & 
\colhead{(days)} & \colhead{(10$^{-13}$ ergs cm$^{-2}$ s$^{-1}$)}}   

\startdata

1215 & $<$0.23 & 4609 & 1.62$\pm$0.06 \\
1448 & 0.07$\pm$0.09 & 4711 & 1.74$\pm$0.07 \\ 
1645 & 0.15$\pm$0.04 & 5038 & 2.59$\pm$0.03 \\ 
1872 & 0.19$\pm$0.04 & 5176 & 2.93$\pm$0.07 \\
2258 & 0.27$\pm$0.05 & 5407 & 3.82$\pm$0.05 \\
2408 & 0.32$\pm$0.07 & 5561 & 4.45$\pm$0.06 \\
2715 & 0.33$\pm$0.11 & 5791 & 5.95$\pm$0.06 \\
3013 & 0.41$\pm$0.06 &      &               \\
\enddata

\tablenotetext{a}{Days since the SN explosion.}

\end{deluxetable}

\clearpage

\end{document}